\documentclass[12pt]{amsart}
\usepackage{geometry}                
\usepackage{setspace}
\usepackage{amssymb}
\title{Mapping an Instanton to Spacetime}
\author{B. E. Eichinger}
\address{Department of Chemistry, University of Washington, Seattle, WA 98118}
\begin{document}
\maketitle
\section*{Abstract}
A mapping from the Lie algebra of the complexified Lorentz group to the $\mathfrak{su}(2)\times\mathfrak{su}(2) \sim\mathfrak{sp}(1)\times\mathfrak{sp}(1)$ part of the algebra the coset space $Sp(2)/[Sp(1)\times Sp(1)]$ is presented. The coset space is shown to be home to the instanton, the curvature form that optimizes the Yang-Mills functional.  Arguments are presented to support the generalization to $Sp(n)/Sp(1)^n$ to yield a self-consistent many-body theory for $n$ particles interacting with one another via fields that reside in the coset space. 
\section*{Introduction}
The relation between the non-compact geometry of spacetime and the compact instanton geometry has been of interest for many years. Instantons, which emerged from Yang-Mills theory, appear to violate classical notions of causality. Yet instantons have become useful tools in quantum theory. Is there more to the instanton than computational convenience?  A deeper understanding of the place of instantons in a general physical structure might answer this question. As a first step toward the answer we will establish the relation between spacetime and instantons beginning with the Lie algebra of the Lorentz group. Known facts about this algebra will be recalled, and compactification of the group with use of the Weyl unitarian trick will show how two copies of $\mathfrak{su}(2)$ emerge. This works because of the underlying group isomorphisms $SO(4)\sim SO(3)\times SO(3)$ and $SO(3)\sim SU(2)$, giving a mapping of algebras $\mathfrak{so}(3,1)\to \mathfrak{su}(2)\times  \mathfrak{su}(2)$.  The map requires that the algebraic field $\mathbb{R}$ has to be extended to the complex field $\mathbb{C}$ by adjunction of $i=\sqrt{-1}$. (Our notation for algebraic fields, rings, groups and their algebras follows standard practice.\cite{Helg,Hump}).

The Yang-Mills functional\cite{YM} is minimized by a curvature two-form over the quaternion algebra $\mathbb{H}$.\cite{Bel,At} The curvature tensor lives in the coset space $Sp(2)/[Sp(1)\times Sp(1)]$; the proof is involved and is left for the second part of this presentation. Accepting the parameterization of the Lie algebra with the coordinates of the coset algebra $\mathfrak{sp}(2)/[\mathfrak{sp}(1)\times\mathfrak{sp}(1)]=:\mathfrak{sp}(2)/\mathfrak{sp}(1)^2$, it will be shown that the two copies of $\mathfrak{sp}(1)\sim\mathfrak{su}(2)$ arise naturally.  The proof of the equivalence of the Lie algebras is straightforward, and as one might guess, it executed by a Wick rotation on the compact side following the Weyl trick on the Lorentz side. But most importantly, it will be shown that just one non-linear operator in the coset algebra generates the two copies of $\mathfrak{sp}(1)$. 

The proof that the instanton appears as the curvature tensor in the coset space requires consideration of some general differential geometry in the setting of a many-body theory. The motivations for constructing a new many-body theory are several. Firstly, relativity is a two-body theory that cannot be easily extended to more objects; a case in point being Bethe-Salpeter theory, which required two time coordinates and is difficult to interpret. The identification of Newtonian gravitational force with curvature in General Relativity and the emergence of the instanton curvature form from particle-level Yang-Mills theory suggests that the $spacetime\leftrightarrow instanton$ mapping is ripe for further analysis.

Some historical precedents have to be suspended if one hopes to achieve a larger view of the relation between the non-compact geometry of relativity and the compact geometry of instantons that goes beyond the map to be discussed. Instantons appear to violate causality; but there is no \emph{a priori} knowledge that causality applies at the  atomic and sub-atomic level of an isolated system. Quantum entanglement also challenges concepts of causality.

The many-body theory is based on the idea that a group $G$, acting transitively on a vector space $V$ that describes the state of a physical system consisting of many parts, can be constructed such that $g\in G$ acts by  $g:V\to gV=\hat V$. There are many physical examples of such an action: (\emph{i})an element of $SO(3)$ acting on an airplane (or sphere) in three space rotates about roll, pitch, and yaw axes, and in so doing generates all orientations of the airplane (the group acts transitively on the sphere); (\emph{ii})the Lorentz group generates all transformations of a spacetime vector with fixed origin. The construction of the many-body theory built on a general group action is the content of the second, larger, section of the paper. There we will circle back to $Sp(2)/[Sp(1)\times Sp(1)]=Sp(2)/Sp(1)^2$ to derive the instanton curvature two-form as a special small system example. The next interesting case is $Sp(3)/Sp(1)^3$, the algebra of which can be understood to consist of three 4D gluons.

\section*{Lorentz and Instanton Algebras}
The mapping from the algebra of the Lorentz group to the algebra of the coset space $Sp(2)/Sp(1)^2$ will be carried out to establish both the relationship between the two, but also to discuss the benefits to be had by generalizing the coset space to higher dimensions.

\subsection*{Algebra of Lorentz Group}
We will choose coordinates $y_i,0\le i\le 3$, where $y_0\sim t$ (units are chosen such that $c=\hbar=1$). The Lorentz algebra is generated by 
\begin{align*}
\mathcal{P}_\alpha:=&(y_0\partial_\alpha+y_\alpha\partial_0),\\
\mathcal{H}_{\beta\gamma}:=&y_\beta\partial_\gamma-y_\gamma\partial_\beta,\; 1\le\alpha,\beta,\gamma\le 3
\end{align*}
with $\partial_i=\partial/\partial y_i$. It is easy to show that $[\mathcal{P}_\alpha,\mathcal{P}_\beta]=\mathcal{H}_{\alpha\beta}, [\mathcal{H}_{\alpha\beta},\mathcal{P}_\gamma]
=\delta_{\beta\gamma}\mathcal{P}_\alpha-\delta_{\alpha\gamma}\mathcal{P}_\beta, \textrm{and}\; [\mathcal{H}_{\alpha\beta},\mathcal{H}_{\beta\gamma}]=\mathcal{H}_{\alpha\gamma}$, so that the general Lie algebra structure $\mathfrak{[p,p]\in h}$, $\mathfrak{[h,p]\in p}$ and $\mathfrak{[h,h]\in h}$ applies.
Now consider the two sets of operators, $L^\pm_\alpha=\pm i \mathcal{P}_\alpha+\mathcal{H}_{\beta\gamma},\{\alpha,\beta,\gamma\}$ cyclic; $i=\sqrt{-1}$, motivated by the Weyl unitarian trick. The commutators are 
\begin{align*}
[L^\pm_\alpha,L^\pm_\beta]=&[\pm i \mathcal{P}_\alpha+ \mathcal{H}_{\beta\gamma},\pm i \mathcal{P}_\beta+ \mathcal{H}_{\gamma\alpha}]\\
=&-2(\pm i\mathcal{P}_\gamma+\mathcal{H}_{\alpha\beta})=-2L^\pm_\gamma\\
[L^+_\alpha,L^-_\beta]=&[i\mathcal{P}_\alpha +\mathcal{H}_{\beta\gamma},-i\mathcal{P}_\beta+ \mathcal{H}_{\gamma\alpha}]\\
=&0
\end{align*}
The conjugate transpose $(L^\pm_\alpha)^*$ of $L^\pm_\alpha$ is defined such that $(L^\pm_\alpha)^*=\mp i\mathcal{P}_\alpha +\mathcal{H}_{\gamma\beta}=-L^\pm_\alpha$ so as to satisfy the requirement that $\mathfrak{g}^*=-\mathfrak{g}$ for $\mathfrak{g}$ the Lie algebra of a unitary group. 

To complete the map to $\mathfrak{su}(1)\times\mathfrak{su}(1)$, define $\eta^\pm_{11}=iL^\pm_3, \eta^\pm_{12}=L^\pm_1+iL^\pm_2$, so that $[\eta^\pm_{11},\eta^\pm_{12}]=-2\eta^\pm_{12}$ and $[\eta^\pm_{11},(\eta^\pm_{12})^*]=2(\eta^\pm_{12})^*$. This splitting of the $SO(4)$ group has been extensively studied by Yang and colleagues.\cite{PSY}

\subsection*{Instanton Algebra}
Turning attention to the instanton half of the relation, first recall that the instanton was found as the curvature two-form that minimizes the Yang-Mills functional,\cite{Bel,At} here written as 
\begin{equation*}
F=(1+\lVert x\rVert^2)^{-2}dx\wedge d\bar x
\end{equation*}
where $x$ is a quaternion (to be defined shortly). The wedge product is in the exterior algebra: $da\wedge db=-db\wedge da$. In the second part of this work it will be shown that this curvature form is one half of the curvature components of the coset space $Sp(2)/Sp(1)^2$. 

The fundamental matrix in the Lie algebra of the coset space is 
\begin{equation*}
 \mathfrak{q}=\left[\begin{matrix} 
      0&q\\
      -\bar q &0
   \end{matrix}\right]\in \mathfrak{sp}(2)/ [\mathfrak{sp}(1)\times\mathfrak{sp}(1)]
 \end{equation*}
where $q$, with conjugate $\bar q$, is an unrestricted quaternion: $q=x_0{\bf 1}+x_1{\bf i}+x_2{\bf j}+x_3{\bf k}; {\bf ijk=-1}$ (details will be provided in the second part). The general parameterization of the Grassmannian algebra has appeared previously\cite{E1} and will be described in detail later, but for now the primary operators are the components of $\mathfrak{p}$ given by (summation convention used in this section)
\begin{equation}\label{instp}
p_{\alpha i}=\partial/\partial\bar\zeta_{\alpha i}+\zeta_{\alpha j}\zeta_{\beta i}\partial/\partial\zeta_{\beta j}.
\end{equation}
Here the $M(2,\mathbb{C})$ matrix representation of a quaternion is used, so that 
\begin{equation*}
q\sim \zeta=\left[\begin{matrix}
z_1&z_2\\
-\bar z_2&\bar z_1\end{matrix}\right]=\left[\begin{matrix}
\zeta_{11}&\zeta_{12}\\
\zeta_{21}&\zeta_{22}\end{matrix}\right]
\end{equation*}
such that $z_1=x_0+ix_3, z_2=x_1+ix_2$.  (The choice of the ${\bf k}$ component on the diagonal of $\zeta$ is consistent with the usual practice in the sciences. Using different fonts for row and column indices of the matrix $\zeta$ is an aid to keep track of the elements in calculations.) Corresponding to the matrix $\zeta$, with components $\zeta_{\alpha i}$, the differential 
$\partial/\partial\zeta_{\alpha i}=\partial_{\alpha i}$ is defined so that $\partial_{\alpha i}\zeta_{\beta j}=\delta_{\alpha\beta}\delta_{ij}$.
Since $\bar\zeta=j\zeta j'$,\cite{BS} it follows that $\partial_{\alpha i}\bar\zeta_{\beta j}=j_{\alpha\beta}j_{ij}$, where 
\begin{equation}\label{jiso}
j=\left[\begin{matrix}
0&1\\
-1&0\end{matrix}\right].
\end{equation}

Now that the tools are in place, the commutator 
\begin{equation*}
[\bar p_{\alpha i},p_{\beta j}]=\delta_{\alpha \beta}H_{ji}+\delta_{ij}h_{\beta \alpha}
\end{equation*}
can be verified. The two $\mathfrak{su}(2)\sim\mathfrak{sp}(1)$ components are 
\begin{align*}
h_{\beta\alpha}=&\zeta_{\beta i}\partial_{\alpha i}-\bar\zeta_{\alpha i}\bar\partial_{\beta i}\\
H_{ji}=&\zeta_{\mu j}\partial_{\mu i}-\bar\zeta_{\mu i}\bar\partial_{\mu j}
\end{align*}
It is not difficult to prove that $[h_{\alpha\beta},H_{ij}]=0$. The components of the two $\mathfrak{sp}(1)$ pieces are written in the complex variables $z_1,z_2$ as  
\begin{align*}
h_{11}=&- h_{22}=(z_1\partial_1-\bar z_1\bar\partial_1+z_2\partial_2-\bar z_2\bar\partial_2)\\
h_{12}=&-\bar h_{21}=2(z_2\bar\partial_1-z_1\bar\partial_2)\\
H_{11}=&-H_{22}=(z_1\partial_1-\bar z_1\bar\partial_1-z_2\partial_2+\bar z_2\bar\partial_2)\\
H_{12}=&-\bar H_{21}=2(z_1\partial_2-\bar z_2\bar\partial_1)
\end{align*}
Note that $h_{11}\;\textrm{and}\;H_{11}$ are pure imaginary. In each sector, we have one commutator to calculate: $[h_{11},h_{12}]=2h_{12}$, since $[\bar h_{11},\bar h_{12}]=[-h_{11},-h_{21}]=[h_{11},h_{21}]=2\bar h_{12}=-2h_{21}$. Similarly, $[H_{11},H_{12}]=2H_{12}$ and $[H_{11},H_{21}]=-2H_{21}$. 

The two $\{\mathfrak{h,H}\}$ operators are next written in terms of the $\mathbb R^4$ components of the quaternion $q=x_0{\bf 1}+x_1{\bf j}+x_2{\bf i}+x_3{\bf k}$ (note that ${\bf j}\sim j$). The differential operators are the usual $\partial/\partial z_1=(1/2)(\hat\partial_0-i\hat\partial_3)$ and $\partial/\partial z_2=(1/2)(\hat\partial_1-i\hat\partial_2)$, where $\hat\partial_i=\partial/\partial x_i$. The definition $m_{ij}=x_i\hat\partial_j-x_j\hat\partial_i, 0\le i\le 3$ will simplify the notation. The operators are 
\begin{align*}
h_{11}=&i[(x_3\hat\partial_0-x_0\hat\partial_3)+(x_2\hat\partial_1-x_1\hat\partial_2)]=i(m_{30}+m_{21})\\
h_{12}=&m_{10}+m_{32}+im_{20}+im_{13}\\
H_{11}=&i(m_{30}-m_{21})\\
H_{12}=&m_{01}+m_{32}+im_{20}+im_{31}
\end{align*}
To check the commutators below, note that 
\begin{equation*}
[m_{ij},m_{kl}]=\delta_{ik}m_{lj}+\delta_{il}m_{jk}+\delta_{jk}m_{il}+\delta_{jl}m_{ki}.
\end{equation*}
The first commutator, written out in detail, is
\begin{align*}
[h_{11},h_{12}]=&i[m_{30}+m_{21},m_{10}+m_{32}+im_{20}+im_{13}]\\
=&i(m_{13}+m_{20}+im_{23}+im_{01}+m_{20}+m_{13}+im_{01}+im_{23})\\
=&2(m_{10}+m_{32}+im_{20}+im_{13})=2h_{12}
\end{align*}
Similarly, $[H_{11},H_{12}]=2H_{12}$.

The Wick rotation $x_0=iy_0$ gives  $m_{0\alpha}\to i\mathcal{P}_\alpha;m_{\alpha 0}\to -i\mathcal{P}_\alpha$, while $x_\alpha=y_\alpha, 1\le\alpha\le 3$, enables the identification $m_{\alpha\beta}=\mathcal{H}_{\alpha\beta}$, so that
\begin{align*}
h_{11}=&i(m_{30}+m_{21})\to i(-i\mathcal{P}_3-\mathcal{H}_{12})=-iL^+_3\\
h_{12}=&-(L^+_1+iL^+_2)\\
H_{11}=&iL^-_3\\
H_{12}=&(L^-_1+iL^-_2)
\end{align*}

Proceeding in the reverse direction, given the generator, eq. (\ref{instp}), one can recover the Lorentz group, and hence special relativity, from a single operator together with a few factors of $\sqrt{-1}$. Is this simply an exercise in executing group isomorphisms, or is there more to it? Special relativity is a two-body theory, designed to be consistent with Maxwell's (macroscopic) equations. The Lie algebra of $Sp(2)$ can be understood to be an active agency that induces transitions between two $\mathbb{C}^2$ vector components (spinors), each of which is acted upon by the $\mathfrak{sp}(1)$ components. But photons induce transitions between two systems -- one losing energy and the other gaining a like amount of energy. We should be thinking of the Lie algebra in the $Sp(2)$ setting as a representation of an \textquoteleft interaction'. Combine this observation with the identification of curvature with force  in general relativity, and curvature with field strength in Yang-Mills, and we have ingredients to  generalize these ideas to more complicated systems. 

For the moment, suppose that the two $\mathfrak{sp}(1)\sim\mathfrak{su}(2)$ components of the algebra of $Sp(2)/Sp(1)^2$ act on two $\mathbb{C}^2$ spinors, and that $q$ and $\bar q$ carry the interaction between the two components. What should we write for three spinors? An extension comes to mind: the next most interesting case will be $Sp(3)/Sp(1)^3$, for which the algebra is spanned by 
\begin{equation*}
\mathfrak{g}(3)=\left[\begin{matrix}
0&c&b\\
-\bar c&0&a\\
-\bar b&-\bar a&0\end{matrix}\right]
\end{equation*}
consisting of three 4D quaternions, $\{a,b,c\}$. Perhaps these can be interpreted as red, green, and blue gluons. This looks too promising to  dismiss as a parlor trick.

\section*{Many-body Theory and Group Theory}
\subsection*{Group and Module}
A many body theory must be capable of treating a composite system as the sum of its parts, with the possibility that the parts may be approximated as independent systems in their own right, especially if the parts are macroscopic. Assume that a system is described by a state vector (a left module) $v$, and that a group $H$ acts so as to transform the state vector: $v\to Hv$. This action may be understood to explore the configuration space of the system. To be physically acceptable, this action should be  subject to restrictions as appropriate to the nature of the system.  A physically acceptable state $Hv$ is as valid a description of the system as is $v$, by assumption. The state of a composite system, consisting of $n$ independent parts is given by $V'=[v_1,v_2,\cdots ,v_n]'\sim v_1\times v_2\times\cdots\times v_n$ (the transpose of $x$ is $x'$), where each of the $v_i$ defines the state of the \emph{i}th system. Corresponding to these separate systems, the group $H:= H_1\times H_2\times \cdots\times H_n$ acts by $H:V\to H_1v_1\times H_2v_2\times\cdots\times H_nv_n$. Of course nothing has been gained by this construction. Since the systems are independent there is no need to discuss more than one of them at a time.

Now allow the systems to interact with one another. To simplify the presentation let $n=2$ (thinking more generally than the instanton$\leftrightarrow$spacetime calculation the extension to any $n>2$ is straightforward). The combined system is given by $V'=[v_1,v_2]'$, but now there is a new group action, $GV$ given by 
\begin{equation*}
GV=\left[\begin{matrix} 
     \hat H_1&G_{12}\\
      G_{21}&\hat H_2
   \end{matrix}\right]\left[\begin{matrix}v_1\\v_2\end{matrix}\right]
\end{equation*}
The assumptions that underlie this action are: (\emph{a}) the components are countable and (\emph{b}) the group $G$ is in the same \textquoteleft family\textquoteright  as $H_i$. With these principles one can build up complex systems and also decompose a system into its parts. The off-diagonal components $G_{ij}\subset G$ are intended to encapsulate the interaction between the two $\{i,j\}$ modules. It is further assumed that $V$ has a finite measure, \emph{i.e.}, $\langle V,V\rangle$ is finite and hence can be normalized regardless of state, implying that $\langle GV,GV\rangle=\langle V,V\rangle$. This requires $G$ to be unitary or orthogonal. To maintain condition (\emph{b}), the diagonal elements are altered: $H_i\to\hat H_i=G_{ii}$ so as to maintain unitarity/orthogonality of $G$. (We anticipate that $H$ and $G$ will correspond to one of the classical algebras: $A_n,B_n,C_n,D_n$\cite{Helg}.)

Now turn this construction around. Suppose we are given a $G$, in any representation, and corresponding $V$. If it happens that $G$ is reducible, then $V$ is decomposable into orthogonal $v_i$, \emph{i.e.}, the system represented by $V$ consists of independent subsystems. On the other hand, for systems that are tightly coupled the elements of $G_{12}$, and its generalization to $G_{ij}$ for $n>2$, are large and the representation is irreducible. This leads to the notion of \emph{almost} reducible, corresponding to small $G_{ij}$ relative to $\hat H_i,\hat H_j$, corresponding to weakly coupled systems.

Let the dimension of $V$ be $m+n$, and suppose that we are given the fundamental representation of $G(m+n)$ along with $H(m),H(n)$, remembering that $G(\cdot)$ and $H(\cdot)$ are in the same family. The action $H(m)\times H(n):V\to H(m)v(m)\times H(n)v(n)$, alters the state of each $v(\cdot)$. One thinks of $H(k)v(k)$ as $H(k)$ varies over its range as the manifold of states accessible to the isolated system. Any point in this manifold can be identified as a fiducial state $v_0(k)$, so identified since  $1\in H(k)$ acts by $1:v(k)\to 1\cdot v_0(k)=v_0(k)$. The coset $G(m+n)/[H(m)\times H(n)]\sim G/H$ is the part of $G$ that alters $v(m)$, resp. $v(n)$, because of the presence of $v(n)$, resp. $v(m)$. 

This construction identifies a principal bundle, with $G$ the total space and $H$ the fiber.\cite{KN} The base space of the bundle is a coset space, most often denoted as $gH$.  The action $H:V\to HV$ shows that $V$ is in the tangent bundle of $G/H$. A small example may be helpful. The tangent space to the 2-sphere, $S^2\simeq SO(3)/SO(2)$, is rotated by the $H\sim SO(2)$ component of $SO(3)$. This rotation is ineffective in that it only  \textquoteleft rotates' the tangent plane and does not  \textquoteleft tilt' it. The action of $SO(2)$ on the tangent space contains no information on the curvature of the sphere; it is the coset $SO(3)/SO(2)$ that acts to \textquoteleft tilt'  the tangent plane, always maintaining it perpendicular to the radial vector, and that is the piece of $SO(3)$ that contains information on the curvature of the sphere. In our case, the curvature residing the the coset codifies the interaction between the subsystems.

\subsection*{Lie Algebra and Physical Fields}
Quantum field theory asserts that matter is constructed from fields. We are discussing a group $G$ that comes equipped with an algebra of vector fields. The vector fields that reside in a Lie algebra may provide a means to represent physical fields. Given a $g\in G$, its eigenvectors exist. Modules can be constructed as linear combinations of eigenvectors of $g$ ($GV$ spans the manifold of states available to the system). In this way the constituents of state vectors are functions of the matrix elements of $G$. But $V$ represents a material state. Simply put, matter states are composed of the fields residing in $G$. This is consistent with the current understanding of quantum field theory.

\subsection*{Ground State and Excitations}
The Lie algebras of the $H(k)$ subgroups induce transitions within each system, as is well known from the action of $\mathfrak{su}(2)$ in elementary quantum mechanics, and which is established more generally by Lie theory.\cite{Helg,Hump,BS} We can anticipate that the fields residing in the algebra of the coset space will induce transitions between systems. The way in which this might happen depends on a detailed examination of representations of the group $G(n)$ having larger dimension than the fundamental representation. Given respective highest weight vectors $v_\mu(m)$ and $v_\nu(n)$, it is conjectured that elements of the Lie algebra of the coset $G/H$ can induce $v_\mu(m)\to v_{\mu\pm1}(m)$ while  $v_\nu(n)\to v_{\nu\mp1}(n)$. This requires that it is possible, when given a representation of $G$ of highest weight $\mu +\nu$, corresponding to $H_\mu$ and $H_\nu$, can be identified and that makes physical sense. The detailed analysis of this conjecture is left to another time.

Now consider the two systems to be in their ground states and further imagine that the combined system, $v(m)\times v(n)$, is in its ground state. The combined system is acted upon by the fundamental representation of $G$. There are no excitations, but $v(m)$ and $v(n)$ still interact with one another \emph{via} the elements of the coset $G/H$. A system with interacting parts, acted upon by the irreducible fundamental representation G, is a ground state system. Clearly this is not restricted to two components; a ground state atom is such a system. (One might entertain thoughts of interacting macroscopic ground state objects.\cite{H}) Before addressing the problem of distinguishing between different types of particles, more fundamentals need to be explained.

\subsection*{Implications for Measurement}
An interlude to discuss what happens in a measurement will cement some of these ideas. Suppose that $v_1$ represents a system, an object to be measured, and $v_2$ an instrument. A measurement is nothing other than a change of state of the instrument that is identifiable as a response that is induced by the system. Clearly if the instrument does not change its state, nothing has happened and no measurement has been made. The instrument changes its state only because there is an interaction between the system and the instrument. At this empirical level we do not need to get involved with wave function collapse or other interpretations based on the quantum theory of isolated systems\cite{Car}; quantum theory is simply not formulated to encompass the interaction between subject and instrument.  

Measurements are not made on isolated systems. In laboratory settings the instrument often includes a means for exciting the system so as to prepare a state that is capable of transmitting an excitation to the detector part of the instrument. (Telescopic observations obviously respond to excitations transferred from the object to the receiver.) The interpretation problem is non-existent, or at least takes an entirely different complexion, if we formulate a theory with a coupling between subject and measuring device, which together constitute a system. A laboratory experiment, considered as an isolated system, consists of both $G$ and $V$ -- fields and material components. 

\subsection*{Geodesics}
Select a subgroup $G_*$ of the group $G$, with corresponding algebra $\mathfrak{g}_*$. A geodesic on the subgroup has the form $\exp(t\mathfrak{g}_*)$ for $t\sim\mathbb{R}$.\cite{JFP,SS} Ultimately $G_*$ can be all of $G$. The upshot of this is that fields flow on geodesics and matter obeys the fields. General Relativity has taught that matter moves in geodesics, but this is not inconsistent with the previous sentence. It may be useful to point out that there are some motions that apparently do not follow geodesic paths. For instance, I doubt that any one of us would claim that going about our daily routine is motion on a geodesic. However, it is not difficult to believe that all of the atomic level fields that create the movements of the day should follow geodesics.  It is clear that \emph{interactions} have to be self-consistent -- they follow well-defined rules as they evolve. (\emph{Field, force, curvature, interaction} -- all are variously used here to describe the same basic principle.) There is room for only one time coordinate to describe geodesics on the group manifold, but one may start the clock at any time to facilitate calculations since $\exp(t\mathfrak{g}_*)V_0=\exp[(t-t_1)\mathfrak{g}_*]\exp(t_1\mathfrak{g}_*)V_0=\exp[(t-t_1)\mathfrak{g}_*]V(t_1)$. This can be put into the form of a differential equation. It is obvious that 
\begin{equation*}
\partial V(t)/\partial t=G(t)\mathfrak{g}V_0=\mathfrak{g}G(t)V_0=\mathfrak{g}V(t).
\end{equation*} 
One should understand $\mathfrak{g}$ as a matrix of first order differential operators that act on the material states. For a subsystem, with the manifold of states $H_iv_i$, the corresponding equation is $\partial v_i(t)/\partial t=\mathfrak{h_i}v_i(t)$. The algebras are many-body operators that are analogous to Hamiltonians.

\subsection*{Algebraic Field $\mathbb{R}$ or $\mathbb{C}$ or Ring $\mathbb {H}$}
Thus far the algebraic ground field or ring  has not been selected, and this can only be done with use of another physical principle. The most direct way to do this is to begin at the bottom. Let our combined system consist of fundamental entities (leptons, quarks, anyons, etc.), each of which cannot be further subdivided by known physical methods. These entities, which are conveniently called particles, are either scalars or they have spin. (Quarks may have additional qualities, \emph{i.e.}, color; these additional variables might be conferred by the group, as intimated above.) Spinor representations of $SO(n)$ have not achieved a fundamental place in quantum field theory, so we will dispense with $\mathbb{R}$ (but $\mathbb{R}$ can also be eliminated by the argument for $\mathbb{C}$ to follow). Consider a Weyl spinor, which is a module $ z=(z_1,z_2)\in\mathbb{C}^2$ that transforms under the action of $SU(2)$. If one hopes to construct higher order representations than the fundamental from Weyl spinors (or Dirac spinors in $\mathbb{C}^4$), tensor products are required. Unfortunately, the irreducible representations of the tensor product $SU(2)\otimes SU(2)$ are modules of dimension 1 and 3. That is, irreducible states constructed from tensor products of fundamental $\mathbb{C}^2$ modules do not maintain the $\mathbb{C}^2$ spinor structure. The reason this is important is that there are composite states, for example protons, that behave as single particles for all but the most extreme conditions. It would be advantageous to use a ground field/ring that maintains this algebraic consistency of spin structure in tensor products so as to represent composite states such as the proton. 

The quaternion ring provides just what is needed. The tensor product of two modules of dimension $k$ having components in the quaternion ring results in a module of dimension $k^2$ (which may be reducible), but the components are still quaternions: $ab\in \mathbb H$ for $a,b\in \mathbb H$. Since a quaternion has four real dimensions, it contains the same information as a $\mathbb{C}^2$ spinor. A quaternion provides an inescapable relation between algebra and the dimension of physical space. 

\subsection*{The Symplectic Group\cite{BS}}
For completeness and to establish notation, a few facts regarding quaternions will be recited so as to establish the structure of the unitary group over the quaternion ring. A quaternion $q$ has components  $q=q_0{\bf 1}+q_1{\bf i}+q_2{\bf j}+q_3{\bf k}$, where, as usual, the basis is ${\bf ii=jj=kk=ijk=-1}$. The conjugate quaternion is $\bar q=q_0{\bf 1}-q_1{\bf i}-q_2{\bf j}-q_3{\bf k}$, such that the norm is $\lVert q\rVert^2=q\bar q=\bar q q=\sum^3_0q^2_\alpha$. (Preference is given to the quaternion basis rather than the Pauli basis; the latter requires the adjunction of a second commuting element, $i=\sqrt{-1}$, which changes both the algebraic simplicity of the quaternion basis and the structure of the symplectic group.) It is convenient to write a quaternion with \textquoteleft scalar' and \textquoteleft vector' parts: $a=a_0+{\bf a}$, with ${\bf a}=a_1{\bf i}+a_2{\bf j}+a_3{\bf k}$. The product $ab$ of two quaternions is $ab=a_0b_0+a_0{\bf b}+b_0{\bf a}+{\bf ab}\sim a_0b_0-{\bf a\cdot b}+a_0{\bf b}+b_0{\bf a}+{\bf a\times b} $, where the second version, with $\mathbb {R}^3$ vector scalar $(\cdot)$ and cross product $(\times)$, are used with care. The quaternion algebra is distributive and additive but not commutative, since $ab-ba=2{\bf a\times b}$, making it a ring and not a field. 

There is an important representation of a quaternion that helps to cement the relation between the algebra and $\mathbb{R}^3$. First, one may write $q=\lVert q\rVert(v_0{\bf 1}+v_1{\bf i}+v_2{\bf j}+v_3{\bf k}=\lVert q\rVert v; v\bar v=1$, so that $v\sim S^3$ is the three sphere. Next, it is well known that a quaternion has a tangent space structure:
\begin{align*}
q=&q_0{\bf 1}+q_1{\bf i}+q_2{\bf j}+q_3{\bf k}=q_0{\bf 1}+{\bf q}=\lVert q\rVert\exp({\bf x})\\
=&\lVert q\rVert[\cos(\lVert x\rVert){\bf 1}+\sin(\lVert x\rVert){\bf u}]; {\bf uu}={\bf x}/\lVert x\rVert=-1.
\end{align*}
Here ${\bf x}=x_1{\bf i}+x_2{\bf j}+x_3{\bf k}$ is a quaternion 3-vector, and ${\bf u}\simeq S^2$; this is  the analogue of $z=re^{i\theta}$. This representation of $q$ identifies the $S^2$ component of the quaternion. The product of two quaternions requires the map $S^3\times S^3\to S^3$, and so on for higher order products, and may require homotopy considerations.  Since the third homotopy group of $S^3$ is $\pi_3(S^3)=\mathbb{Z}$, it is possible that different classes of particles belong to different homotopy classes. The study of Hopf fibrations of $S^3$ should prove interesting in application to particle physics.\cite{U} The abstract $S^3$ part of a particle state can be mapped to ${\bf x}\sim\mathbb{R}^3\cup\infty$ by the antipodal projection ${\bf x}={\bf q}/q_0$.

The unitary group of dimension $n$ over the quaternion ring $\mathbb{H}$, temporarily denoted as $U(n,\mathbb{H})$, is defined such that for $\mathcal{U}\in U(n,\mathbb{H}); \mathcal{U^*U=UU^*}=1$, where $\mathcal {U}^*$ is the transpose conjugate of $\mathcal{U}$: each element $\upsilon_{ji}\in\mathcal{U}^*$ is the conjugate of $\upsilon_{ij}\in\mathcal{U}$. This group is better known as $Sp(n)$. There is a map\cite{V}: $U(n,\mathbb {H})\to Sp(2n,\mathbb C), n>1$, which will not be used here but which might be useful, should the noncommutative ring $\mathbb {H}$ be troublesome for a particular calculation. Here the $\mathbb{H}\to M(2,\mathbb {C})$ representation is preferred, where $M(2,\mathbb {C})$ is a two-by-two matrix; it makes essential use of the well-known isomorphism between the Lie algebras $\mathfrak{sp}(1)\simeq\mathfrak{su}(2)$. Define the matrix representation of a quaternion $q$ as 
\begin{equation}\label{Mofq}
q\simeq\left[\begin{matrix} 
      z_1&z_2\\
      -\bar z_2 &\bar z_1
   \end{matrix}\right]=\zeta =\left[\begin{matrix} 
      \zeta_{11}&\zeta_{12}\\
     \zeta_{21} &\zeta_{22}
   \end{matrix}\right]
\end{equation}
An easy calculation shows that $\bar q\simeq\zeta^*$. The complex conjugate of $\zeta$ is $\bar\zeta$ and $\zeta'$ is the transpose; the two combined are conjugate transpose: $\zeta^*:=\bar\zeta'$.

\subsection*{Cosets of $Sp(n)$}
The $G/H$ construction described above may now be written, for the simplest case of two systems as $Sp(n)/[Sp(k)\times Sp(n-k)]$. This is a coset space that is also identified as a Grassmann manifold over the quaternions. Clearly the coset construction can be extended to $Sp(n)/[Sp(k_1)\times Sp(k_2)\times\cdots Sp(k_m)]; \sum k_i=n$. Now consider this taken to the level of elementary particles, \emph{i.e.}, $Sp(n)/[Sp(1)\times Sp(1)\times\cdots\times Sp(1)]=:Sp(n)/Sp(1)^n$. What is the action of $Sp(1)^n$? To understand this, return to the spinor $z=(z_1,z_2)$. The norm of $z$ is invariant to the action of $U(2,\mathbb{C})$. However, the abelian group $\exp(i\theta)1$, which is $U(2,\mathbb{C})/SU(2,\mathbb{C})$, simply multiplies each component by a phase factor: $z_i\to e^{i\theta} z_i; i=1,2$: the phase factor is \textquoteleft ineffective' in altering the state of the spinor, so that $SU(2,\mathbb{C})$ is the \textquoteleft effective' part of $U(2,\mathbb{C})$. A similar interpretation applies to $Sp(1)^n$; its action simply rotates each component: $v_i\to Sp(1)v_i$, independent of the other components of $V$ and does not change the state of $V$. One can think of $Sp(1)$ as the quaternion unit vector. 

To return to the idea of fields creating particles, consider the primeval state $V_0=(1,0,0,\cdots,0)'$ such that for $g\in G:gV_0=V$. But $V_0$ is invariant to the action of $Sp(n-1)$, enabling one to identify $V\simeq Sp(n)/Sp(n-1)$ as a coset space: $V=S^{4n-1}$. There are two coset spaces at work: one for matter states and the other for interactions.

A remark concerning cosets in general will provide a tool that is essential for calculations. The coset $G/H$ has both left and right inverses and contains an identity, but it is not a group because it does not have an appropriate product. To see this, let $x\in G/H$. Then an element of the coset is $xH$. Restricting $G$ and $H$ to be unitary/orthogonal so that the inverse to $\mathcal{G}\in G$ is the transpose conjugate $\mathcal{G}^*$, it is easy to see that $1=xH(xH)^*=xHH^*x^*=xx^*$ gives the right inverse. For the left inverse, $(xH)^*(xH)=H^*x^*xH$; multiply on the left by $H$ and on the right by $H^*$ to get $x^*x=(H^*)^{-1}H^{-1}=(HH^*)^{-1}=1$, which proves there is a left inverse. However, $xHyH\ne xyH$; to map one element of a coset to another requires the action of all of $G$ in general. 

\subsection*{Curvature Tensors}
The action $g:V\to gV$ is the fundamental process that changes the state of the module $V$. Restricting $g$ to a unitary/orthogonal group, it follows that the inner product, $\langle V,V\rangle$ is invariant to the $g$ action. Given a fixed (fiducial) state $V_0$ it follows that $V=gV_0$ since $G$ acts transitively. An infinitesimal change in $V$ is given by the (exterior) derivative $dV=dgV_0=(dg)g^*V=\omega V$, which is the first Maurer-Cartan form. The skew-symmetric differential form $\omega$ contains all of the essential information to understand the geometry of the theory.  

To show how this develops, consider the metric on the space of matrices. A natural metric for a matrix $g\in GL(n,\mathbb{K})$, is $ds^2=Tr(dg^*dg)=Tr(dgdg^*)$, where $\mathbb{K}$ is $\mathbb{R,C,H}$, and $GL(n,\cdot)$ is the general linear group. (The trace of a matrix product over $\mathbb{H}$ is invariant to cyclic permutations if the trace is real.) Restricting $g$ to a unitary/orthogonal group $\mathcal{U}(n,\mathbb{K})$, reduces this to $ds^2=Tr(dgg^*gdg^*)=Tr(\omega\omega^*)$. The important point is that $\omega$ is a right invariant differential form: $g\to gk$ for fixed $k\in \mathcal{U}(n,\mathbb{K})$ gives $dgkk^*g^*=\omega$. 

The exterior derivative of $dV$ is $d^2V=0=d\omega V-\omega\wedge dV=(d\omega-\omega\wedge\omega)V\Rightarrow d\omega-\omega\wedge\omega=0$, which is the second Maurer-Cartan (MC2) form. For a given $k_\mu$ partition of $n$, let the index set of the blocks be denoted by capital letters, so that the components of the MC2 form can be written as 
\begin{align*}
d\omega_{IJ}-\sum_K\omega_{IK}\wedge\omega_{KJ}=&0\\
d\omega_{II}-\omega_{II}\wedge\omega_{II}=&\sum_{K\ne I}\omega_{IK}\wedge\omega_{KI}
\end{align*}
which identifies the curvature two-forms\cite{Cartan}
\begin{equation}\label{curv}
\Omega_{II}=-\sum_{K\ne I}\omega_{IK}\wedge\omega_{KI}=\sum_{K\ne I}\omega_{IK}\wedge\bar\omega_{IK}.
\end{equation}
This identification of the curvature forms $\Omega_{II}$ follows Chern.\cite{Chern46} The signs in MC2 are determined by the choice of left action of the group. If one chooses a right action, $\omega_r=g^*dg$ and $d\omega_r+\omega_r\wedge\omega_r=0$; the two forms are related by a similarity transformation, or conjugacy, $\omega_r=g^*\omega g$.

For the given $k_\mu$-partition, consider a change of basis for $h\in H,V\to hV=\hat V$, which corresponds to a different selection of cross-section of $G/H$.\cite{Chern95}. The associated connection form is $d\hat V=\hat\omega\hat V$, and the exterior derivative of $\hat V$ is 
\begin{equation*}
d\hat V=\hat\omega\hat V=dhV+hdV=dhh^*\hat V+h\omega h^*\hat V
\end{equation*}
giving $\hat\omega=h\omega h^*+dhh^*$.  This is familiar as a gauge transformation. The exterior derivative of this equation, with subtraction of $\hat\omega\wedge\hat\omega$, gives 
\begin {equation*}\label{chg}
d\hat\omega-\hat\omega\wedge\hat\omega = h(d\omega-\omega\wedge\omega) h^*
\end{equation*}
from which it can be seen that $\hat\Omega_i=h_i\Omega_ih^*_i$, demonstrating the tensor character of $\Omega_i$ and invariance to a gauge transformation. Chern\cite{Chern95} develops a string of invariants $\mathcal{I}(i)_m=\Omega_i\wedge\Omega_i\wedge\cdots\wedge\Omega_i, m\; \textrm{terms},$ which may prove useful. On of these is considered later. 

For the simplest cases, $n=2,3$, it will be useful and instructive to write out the two and three components of the curvature 2-forms. For $Sp(2)/Sp(1)^2$ we have 
\begin{equation*}
\Omega_1=\Omega_{11}=\omega_{12}\wedge\bar\omega_{12}\textrm{ and }\Omega_2=\bar\omega_{12}\wedge\omega_{12}=-\Omega_1.
\end{equation*}
We will return to these equations later to show how they are related to the instanton. A delightful symmetry is revealed for $Sp(3)/Sp(1)^3$\cite{E2}, for which 
\begin{align*}
\Omega_1=&\omega_{12}\wedge\bar\omega_{12}+\omega_{13}\wedge\bar\omega_{13},\\
\Omega_2=&\bar\omega_{12}\wedge\omega_{12}+\omega_{23}\wedge\bar\omega_{23},\\
\Omega_3=&\bar\omega_{13}\wedge\omega_{13}+\bar\omega_{23}\wedge\omega_{23}.
\end{align*}
The $n=3$ coset space is identified as the 12 dimensional Wallach space.\cite{W1} A deep and potentially interesting aspect of this space is that the curvature may develop negative components,\cite{W2} which is a subject for another time.

\section*{Grassmann Geometry}
Rather than continue with the general development of $Sp(n)/Sp(1)^n$, which is not trivial, we will turn attention to the Grassmann manifold where explicit calculations are easier; the case $Sp(n+1)/[Sp(1)\times Sp(n)]$ will also be most interesting for understanding the relation between many-body theory and traditional two-body theory. A matrix $g\in Sp(k+n)$ can be partitioned 
\begin{equation}\label{partition}
g=\left[\begin{matrix}
A(k\times k)&B(k\times n)\\
C(n\times k)&D(n\times n)\end{matrix}\right]=\left[\begin{matrix}
1&X\\
-X^*&1\end{matrix}\right]\left[\begin{matrix}
A&0\\
0&D\end{matrix}\right]
\end{equation} 
where $X=BD^{-1}=-(A^*)^{-1}C^*$, with the latter version resulting from $A^*B+C^*D=0$. It is easy to show that $AA^*=(1+XX^*)^{-1};DD^*=(1+X^*X)^{-1}$. The $k\times n$ matrix $X$ is isomorphic to a Grassmann manifold. 

\subsection*{Metric and Laplacian}
A calculation shows that the metric\cite{E1,Hua} is given by
\begin{equation*}
ds^2=\textrm{Tr}[(1+XX^*)^{-1}dX(1+X^*X)^{-1}dX^*]=\textrm{Tr}[R^{-1}dXS^{-1}dX^*],
\end{equation*}
where $R=1+XX^*, S=1+X^*X$; the metric is left invariant: $\hat g:g\to \hat gg$. One begins by observing that $\hat g: X\to (\hat AX+\hat B)(\hat CX+\hat D)^{-1}$, which is a linear fractional transformation and the start of a lengthy calculation. The similarity to a K\"ahler metric is evident.

The Laplace-Beltrami operator, $\mathcal{L}$, is dual to the metric and is written in components (summation convention) as 
\begin{equation}\label{LB}
\mathcal{L}=(1+X^*X)_{ij}(1+XX^*)_{\alpha\beta}\frac{\partial^2}{\partial x_{\alpha j}\partial\bar x_{\beta i}}
\end{equation}
so as to place the differential operators to the right of the functions of coordinates. Using an obvious symbolic notation, this is more simply written as $\mathcal{L}=\textrm{Tr}(R\bar\partial S\partial')$, with the understanding that $\partial$ does not act on coordinates to the right. It is claimed that the quadratic terms in this operator are the source of the nonlinear terms in the standard model, which will become clearer after displaying the generators of the Lie algebra. Proof that this theory can be mapped to the standard model, or the other way around, will be a heroic task.

\subsection*{Lie Algebra}
Given a $g\in G$, the corresponding algebra $\mathfrak{g}$ is $G\simeq\exp\mathfrak{g}$. The Grassmannian $Sp(k+n)/[Sp(k)\times Sp(n)]$ has a coset algebra and coset given by
\begin{equation*}
\mathfrak{g}\sim\left[\begin{matrix}
0&Q\\
-Q^*&0\end{matrix}\right]
\end{equation*}
\begin{equation*}
\exp(\mathfrak{g})=\left[\begin{matrix}
\cos(QQ^*)^{1/2}&(QQ^*)^{-1/2}\sin(QQ^*)^{1/2}Q\\
-Q^*(QQ^*)^{-1/2}\sin(QQ^*)^{1/2}&\cos(Q^*Q)^{1/2}\end{matrix}\right]\label{eg}
\end{equation*}
respectively, where the trigonometric functions of the matrices $(QQ^*)^{1/2}$ and  $(Q^*Q)^{1/2}$ are defined by their Taylor series expansions.  This enables one to identify
\begin{equation*}
X=(QQ^*)^{-1/2}\sin(QQ^*)^{1/2}Q(\cos(Q^*Q)^{1/2})^{-1}
\end{equation*}
 if desirable.
 
 A geodesic on $g\in G$ is given by $g=\exp(t\mathfrak{x})$, with 
 \begin{equation*}
 t\mathfrak{x}=\left[\begin{matrix}
 th(k)&tx\\
 -tx^*&th(n)\end{matrix}\right]=\left[\begin{matrix}
0&tx\\
 -tx^*&0\end{matrix}\right]+\left[\begin{matrix}
 th(k)&0\\
 0&th(n)\end{matrix}\right].
 \end{equation*}
This enables one to see that 
\begin{equation*}
g(t)=c(t)H(t)
\end{equation*}
where $c(t)$ is the coset. This makes it possible to focus on the time dependence of the coset, since the $H(t)$ part factors out from $X(t)$, and this generalizes to more than two components in the flag manifold.

A matrix $g\in Sp(n)$ has two instantiations, one with quaternion elements and the other with $\zeta\in M(2,\mathbb{C})$ matrix elements. We will freely switch from one to the other, but will tend to use the quaternion version for general discussion and the $\zeta$ version for calculations. The infinitesimal generators for the \textquoteleft active' piece of the Lie algebra for a group are built from the coset coordinates. To keep track of components when doing calculations it is convenient to use different fonts for the row and column indices of $\zeta=(\zeta_{\alpha a})$, so that $(\zeta_{\alpha a})'=\zeta_{a\alpha}$. A $k\times n$ matrix consisting of components of the form eq. (\ref{Mofq}) has $2k$ rows and $2n$ columns, with the $M(2,\mathbb {C})$ retaining their natural order. In evaluating commutators, the $j$ isomorphism, eq. (\ref{jiso}) is essential.

The author has shown\cite{E1} that the three \textquoteleft effective' parts of $\mathfrak{sp}(n)$ are given by (summation convention)
\begin{align}\label{gen}
p_{\alpha a}=&\bar\zeta_{\alpha a}+\zeta_{\alpha b}\zeta_{\beta a}\partial_{\beta b}\\
h_{\alpha\beta}=&\zeta_{\alpha a}\partial_{\beta a}-\bar\zeta_{\beta a}\bar\partial_{\alpha a}\nonumber\\
H_{ab}=&\zeta_{\alpha a}\partial_{\alpha b}-\bar\zeta_{\alpha b}\bar\partial_{\alpha a}\nonumber
\end{align}
The general structure: $\mathfrak{[p,p]\in h, [h,p]\in p, [h,h]\in h}$, can be verified by the reader. Clearly both $h$ and $H$ are skew-symmetric, and $p^*_{a\alpha}= -\bar p_{a \alpha}$. A shorthand notation is useful: the second of these equations can be written as $h=\zeta\partial'-(\zeta\partial')^*$, and with the use of the $j$ conjugation, this becomes $h=\zeta\partial'-(j\zeta\partial'j')'$, which enables one to keep track of the components when evaluating commutators. To complete the set, $H=\zeta'\partial-(\zeta'\partial)^*, p=\bar\partial+\zeta(\zeta'\partial)'=(1+\zeta\zeta^*)\bar\partial+\zeta H'$. The complete algebra of $Sp(n)$ includes contributions from the $h_k\times h_n$ fiber components of $G/H$; these components do not transfer excitations between subsystems, but each transfers excitations within the respective subsystems. 

The homogeneous generators, $h$ and $H$, derive from the primary field given by eq. (\ref{gen}). These \textquoteleft effective' parts of the Lie algebra of $\mathfrak{sp}(k+n)$ act on $v_k$ and $v_n$, respectively. To emphasize the potential relation os eq. (\ref{gen}) to the standard model of particle physics, consider $Sp(n+1)/(Sp(1)\times Sp(n)$. The matrix  of the coset variables is the $2\times 2n$ matrix $Q=(\zeta_1,\zeta_2,\cdots,\zeta_n)$, The $p_{\alpha i}$ component can be written\cite{RG}
\begin{align*}
p_{\alpha i}=&\bar\partial_{\alpha i}+\sum^n_{j=1}\sum^2_{\beta=1}\zeta_{\alpha j}\zeta_{\beta i}\partial_{\beta j}\\
=&\bar\partial_{\alpha i}+\zeta_{\alpha i}\zeta_{\beta i}\partial_{\beta i}+\sum^n_{j\ne i}\sum^2_{\beta=1}\zeta_{\alpha j}\zeta_{\beta i}\partial_{\beta j}.
\end{align*}
This can be mapped to spacetime by the procedures in the first part should that be desirable. Given that the standard model represents field components as quadratic terms, this equation provides a possible route to a deeper understanding of the interactions of a particle with its neighboring particles. The generalization to two or more particles and their interactions with their environment is evident in eq. (\ref{gen}). 

\subsection*{Specializing to $Sp(2)/Sp(1)^2$}

The metric and LB operators for this case are (summation convention)
\begin{align*}
ds^2=&(1+\lVert x\rVert^2)^{-2}dx_{\alpha i}d\bar x_{\alpha i},\\
\mathcal{L}=&(1+\lVert x\rVert)^2)^2\frac{\partial^2}{\partial x_{\alpha i}\partial\bar x_{\alpha i}},
\end{align*}
the latter from eq. (\ref{LB}). At this point it is tempting to make the following approximation. For small $\lVert x\rVert$  the Laplace-Beltrami operator is just $(1/2)\sum^3_0\partial^2/\partial^2 x_i$, and given the mapping $y_0=\sqrt{-1}x_0\sim t$, one might consider a stationary harmonic function that is the solution of $\nabla^2 \phi=0$, for example. This naive route to the Coulomb potential is merely illustrative, and is not pursued further.

 The curvature two-forms from eq. (\ref{curv}) are 
\begin{align*}
\Omega_1=&\omega_{12}\wedge\bar\omega_{12}\\
\Omega_2=&\bar\omega_{12}\wedge\omega_{12}\\
\end{align*}
An explicit calculation based on the Grassmann coordinates will show how the Atiyah\cite{At} two-form is obtained. We have 
\begin{equation*}
dgg^*=\left[\begin{matrix}
dAA^*+dBB^*&dAC^*+dBD^*\\
dCA^*+dDB^*&dCC^*+dDD^*\end{matrix}\right]=\left[\begin{matrix}
\omega_{11}&\omega_{12}\\
\omega_{21}&\omega_{22}\end{matrix}\right]
\end{equation*}
Since $X=BD^{-1}=-(A^*)^{-1}C^*$ it follows that $dXD+XdD=dB,A^*dX+dA^*X=-dC^*$, giving
\begin{align*}
\omega_{12}=&dAC^*+dBD^*=dA(-A^*X)+(dXD+XdD)D^*\\
=&-\delta AX+X\delta D+dXDD^*\\
=&-\delta AX+X\delta D+dXS^{-1}
\end{align*}
where $\delta A=(dA)A^*, \delta D=(dD)D^*$. Similarly, 
\begin{align*}
\omega_{21}=&dCA^*+dDB^*=-(dX^*A+X^*dA)A^*+dDD^*X^*\\
=&\delta DX^*-X^*\delta A-dX^*R^{-1}
\end{align*}
As before, $AA^*=(1+XX^*)^{-1}=R^{-1};DD^*=(1+X^*X)^{-1}=S^{-1}$. For the $n=2$ case these forms can be simplified since $A=D=\cos(\lVert x\rVert)$, so that $\delta A=\delta D$ are scalars (more precisely, multiples of the ${\bf 1}$ element of the quaternion basis), giving $\omega_{12}=dx(1+\lVert x\rVert^2)^{-1}, \omega_{21}=-d\bar x(1+\lVert x\rVert^2)^{-1}$. We finally have
\begin{align*}
\Omega_1=&(1+\lVert x\rVert^2)^{-2}dx\wedge d\bar x,\\
\Omega_2=&(1+\lVert x\rVert^2)^{-2}d\bar x\wedge dx.
\end{align*}
These are anti-self-dual and self-dual curvature forms, respectively, as was first shown by Atiyah.\cite{At} Note that $\Omega_2=-\Omega_1$. As promised, the coset space $Sp(2)/Sp(1)^2$ contains the instanton as well as the Lie algebra that maps to the $SO(3,1)$ algebra. 

To complete this section, a Chern invariant for $Sp(2)/Sp(1)^2$ will be calculated; it is given by
\begin{align*}
I=&\int{|\Omega_i\wedge\Omega_i}|=\int(1+\lVert x\rVert^2)^{-4}|dx\wedge d\bar x\wedge dx\wedge d\bar x|\\
=&24\int(1+\lVert x\rVert^2)^{-4}|dx_0\wedge dx_1\wedge dx_2\wedge dx_3|\\
=&48\pi^2\int^\infty_0(1+y^2)^{-4}y^3dy=48\pi\int^\infty_0\frac{\sinh^3t}{\cosh^7t}dt\\
=&48\pi^2\int^\infty_1\left(\frac{1}{s^5}-\frac{1}{s^7}\right)ds=4\pi^2
\end{align*}
Whether invariants such as this will be important in particle theory remains to be seen. This exercise is simply to illustrate that at least one invariant is easily calculable; since the spaces involved are compact there is every reason to believe that higher order invariants for $Sp(n)/Sp(1)^n,n>2$ will not present problems.

Returning to Hopf fibrations, the two particle state module $(v_1,v_2)$ has the geometry $S^7$, so the fibration $S^3\hookrightarrow S^7\to S^4$ will be applicable. In addition to these fibrations, there is a wealth of four dimensional theory that may prove useful.

\section*{Conclusion}
 
Given the Lie algebra of $Sp(n)/Sp(1)^n$, one may choose any pair of conjugate quaternions, $q_{ij}$ and $q_{ji}=-\bar q_{ij}$, and treat these in isolation to give the special relativity geometry, just as we have done for the $n=2$ case. However, this procedure cannot be extended to more than one pair of quaternions, as an attempt to force more than one pair into special relativity will result in multiple time coordinates, again creating a problem for interpretation. A self-consistent many-body theory can only be constructed with instantons.

The mapping between the Lie algebras that has been used here to illustrate the intimate connection between the hyperbolic geometry of relativity and the spherical geometry of instantons is not the only path to this equivalence. It is illuminating to cite two others. Humphreys\cite{Hump} analyzes Lie algebras in general by showing that every Lie algebra contains multiple copies of $\mathfrak{sl}(2,\mathbb{K})$, where $\mathbb{K}$ is a rank-one algebra (this is true because every root has a reflective partner). Since $SL(2,\mathbb{C})/\mathbb{Z}_2$ is homomorphic to $SO^+(3,1)$, every Lie algebra contains many copies of the non-compact Lorentz algebra. The $\{q_{ij},-\bar q_{ij}\}$ pair is but one example of this. Yet another way to relate $y^2_0-{\bf y\cdot y}=\sigma^2$ to $x^2_0+{\bf x\cdot x}=\rho^2$, where ${\bf y,x}\in\mathbb{R}^3$, is via the stereographic projections 
\begin{equation*}
{\bf r\cdot r}=\frac{\bf y\cdot y}{(|y_0|+\sigma)^2}=\frac{|y_0|-\sigma}{|y_0|+\sigma}=\frac{\bf x\cdot x}{(|x_0|+\rho)^2}=\frac{\rho-|x_0|}{\rho+|x_0|}<1
\end{equation*}
of the hyperbola and sphere into the 3D ball. Which of these several routes to mapping between particle theory and this compact many-body theory remains to be seen. 

Returning to entanglements, an isolated laboratory experiment should be understood as a material apparatus together with the interactions between its parts, and the combination exists in an irreducible quantum state. The detection of a state of a subsystem is necessarily matched by a compensating change in the complementary subsystem, whether observed or not. For example, a spin zero system might be separated into two components -- if one is found with spin up the other must have spin down; keep in mind that each observation is recorded by a change of state of material components (detectors). Since no system is truly isolated, perturbations can be expected from interactions between the experimental system and its surroundings. These interactions are embodied in the coset $G/H$, and may be regarded as perturbations in the sense that in dividing a system from its surroundings one is dealing with an \emph{almost} reducible representation.

There exists a multitude of directions that might be pursued to put flesh on this outline of many-body theory. One of these directions is provided by the coset $Sp(3)/Sp(1)^3$, which has immediate applications to baryon structure. The corresponding (isolated and normalized) matter state is the eleven-sphere, $S^{11}=S^{4n-1}$, and the gluon field is 12 dimensional. The gluons are required to explain high energy experiments, yet this picture coexists with the notion that at low energies composite states are derived from tensor products of the fundamental representation (including linear combinations of tensor product components). A ground state system consisting of interacting parts does not exchange excitations by definition; the fundamental representation of the group acts on the ground state. There is obviously much work that needs to be done to make this theory useful -- it will require expertise from both mathematical and physical sciences.

\end{document}